\documentclass{llncs}

\usepackage{color}
\usepackage{moreverb}
\usepackage{subfigure}
\usepackage{graphicx}
\usepackage{floatflt}

\newcommand{\moise}[0]{$\mathcal{M}\textsc{oise}^+$}
\newcommand{\smoise}[0]{$\mathcal{S}$-\moise}
\newcommand{\jmoise}[0]{$\mathcal{J}$-\moise}
\newcommand{\jason}[0]{\textbf{\emph{Jason}}}

\newcommand{\us}{\underline{\hspace*{0.2cm}}}

\begin{document}

\title{An Investigation of the Advantages of Organization-Centered Multi-Agent Systems}

\author{Andreas Schmidt Jensen}

\institute{Department of Informatics and Mathematical Modelling \\ Technical University of Denmark \\ Richard Petersens Plads, Building 321, DK-2800 Kongens Lyngby, Denmark}

\maketitle \pagestyle{plain} \thispagestyle{plain}

\medskip

\begin{abstract}
Whereas classical multi-agent systems have the \emph{agent} in center, there have recently been a development towards focusing more on the \emph{organization} of the system. This allows the designer to focus on \emph{what} the system goals are, without considering \emph{how} the goals should be fulfilled. This paper investigates whether taking this approach has any clear advantages to the classical way of implementing multi-agent systems. The investigation is done by implementing each type of system in the same environment in order to realize what advantages and disadvantages each approach has.
\end{abstract}

\medskip

\section{Introduction}
Within the area of multi-agent systems there has recently been a development towards making the organization of such systems explicit \cite{Ferber+2004,Hannoun+2000,Hubner+2002}. However, while \cite{Ferber+2004} lists some drawbacks of classical (agent-centered) multi-agent systems, the actual advantages of making the organization explicit has not been thoroughly investigated. 

This paper summarizes our work with such investigation of the organization of multi-agent systems. The investigation was conducted by implementing two systems: a classical (agent-centered) (ACMAS) and an organization-centered (OCMAS).

In classical multi-agent systems the agent is in focus. The programmer developing the agents is able to decide what the agents can do and \emph{how} the choose to do it. In an OCMAS we are more concerned with the organization; i.e. the structure of the multi-agent system. Naturally all multi-agent systems have a structure, but it is most often implicitly defined by the agents and their relations.

By explicitly defining the organization it is possible to focus on \emph{what} the agents should do without at the same time deciding \emph{how} they should do so. In other words, the organization makes it possible to create the structure of the system without specifying details about the implementation.

\section{The Setup}
The two types of systems have been compared using a team-based version of the well-known game Bomberman. However, the nature of an implementation of intelligent agents does not guarantee a certain quality and a comparison based on the overall performance of a team of agents may not be adequate; the results may merely be caused by better or worse strategies. 

Since the two approaches are quite different in many ways, it seems more natural to employ other measures of comparison. The comparison of ACMAS and OCMAS is therefore based on the following measures:
\begin{itemize}
	\item Structure of the source code
	\item Development speed
	\item Performance
	\item Error handling
	\item Debugging
	\item Complexity of the scenario
	\item Number of intelligent agents
\end{itemize}

The original Bomberman game consists of five key elements: bombs, boxes, solid obstacles, exitways and power-up panels. Whereas boxes are destructible, solid obstacles are not. This means that Bomberman will always be able to take cover behind solid obstacles. Most boxes must be destroyed since they hide both power-ups and exitways. Exitways are what Bomberman must find to be able to complete a level. A power-up can be used to enhance bomberman's abilities and bombs. In the beginning, his bombs are weak, but by using power-ups he will be able to drop several stronger bombs at a time. 

In a multi-agent context, the enemies could be considered a team of agents (i.e. a multi-agent system) with the general purpose of stopping Bomberman from escaping and by implementing the enemies using each of the approaches we could perform the comparison. However, to be able to experiment with the cooperative aspects of intelligent agents we instead propose an altered version of Bomberman in which two teams attempt to eliminate each other:

\begin{definition}[Team-based Bomberman]\label{def-scenario}
The multi-agent system is similar to Bomberman. It consists of two teams fighting against each other. Each team consists of at least two ``bombermen'' (or agents). The teams are situated in a maze-like environment consisting of solid obstacles and boxes. An agent can place bombs which at some point will explode. An agent dies when he is hit by an explosion. Explosions will also destroy boxes. A team wins when all players from the other team have been eliminated. 
\end{definition}

This version of Bomberman consists of some of the same key elements as the original game: a maze, destructible and indestructible obstacles, and bombs. This should allow the agents to employ the strategies intended for the game, while at the same time competing in teams. This fact creates a new aspect of the game, since a group of agents potentially is able to trap enemies by placing bombs strategically.

The concepts of exitways and power-ups have been excluded in this version. Exitways have been removed, since the overall goal of ``getting to the surface'' is no longer relevant (as the goal instead is to eliminate the other team). Power-ups are not included to avoid making the overall system too complex since the intention is not to make a perfect implementation of Bomberman; rather is it to compare and discuss two different approaches to implementing multi-agent systems. In this case we believe a simple, yet strategically challenging system will be adequate. 

\subsection{\jason}
The implementation of the ACMAS is done using \jason, \emph{``a Java-based interpreter for an extended version of AgentSpeak''}\footnote{\url{http://jason.sourceforge.net/}}. We provide an overview of the interpreter by introducing how to program multi-agent system using it, however we will not go into details with all parts of the system. The overview should give a foundation for building simple systems using \jason. A thorough description of \jason\ is found in \cite{Bordini+2007}.

The language of \jason, AgentSpeak, is a Prolog-like logic programming language. AgentSpeak allows the developer to create a \emph{plan library} for the agent. A plan in AgentSpeak is basically of the form 
\[
\texttt{+triggering\_event} : \texttt{context} \texttt{ <- } \texttt{body}.
\]

Roughly speaking, if an event matches a trigger, the context is matched with the current state of the agent. If the context matches the current state, the body is executed; otherwise the engine continues to match contexts of plans with the same trigger. If no plan is applicable, the event fails.

The fact that AgentSpeak is a logic programming language allows one to easily transfer specifications written in logic formulas of a multi-agent system to an implementation written in \jason. For instance, part of a plan for a vacuum cleaner agent \cite{Russell+2003} is shown below:
\[
\texttt{+!cleaning} : \texttt{in(X,Y) \& dirt(X,Y)} \texttt{ <- } \texttt{do(suck).}
\]

The plan is triggered by the goal \texttt{!cleaning}, so if the vacuum cleaner is in a ``cleaning state'', this triggering event would be applicable. The context specifies that this plan is relevant if the agent currently is somewhere in the environment which is dirty. If the context can be unified with data from the database of the agent, it will perform the body, which in this case means that it will perform the action \texttt{do(suck)}. However, as mentioned it is possible to have several plans for the same triggering event if those plans have different contexts:
\[
\texttt{+!cleaning} : \texttt{in(X,Y) \& dirt(X+1,Y)} \texttt{ <- } \texttt{do(right).}
\]

This plan will then be applicable if the agent has perceived dirt in an area to the right of its current area. In that case, it will perform the action \texttt{do(right)}.

\subsection{The \moise\ organizational model}
The implementation of the OCMAS is based on the \moise\ organizational model, in which it is possible to create a structural, functional and deontic specification of an organization. The organizational model has been combined with \jason\ in the middleware called \jmoise.

\moise\ is an organizational model for multi-agent systems which makes it possible to specify the organization in a MAS structurally, functionally and deontically. The model takes an organizations-centered approach, meaning that an organization will exist \emph{a priori} (created at design-time) and the agents ought to follow it \cite{Hubner+2002}. 

\paragraph{Structural Specification:}
\moise\ uses the concepts of roles, role relations and groups in the structural specification of an organization. Each agent plays one or more roles. The roles are related by links, which specify how agents are acquainted and can communicate. In order to further structure the organization, the agents can join different groups depending on the roles they play.

\paragraph{Functional Specification:}
The functional specification consists of a \emph{goal decomposition tree}, known as a Social Scheme (SCH), where the root is the goal of the SCH and each node is a sub-goal that can be delegated to different agents. In \cite{Hubner+2002} three operators are defined for decomposing a goal into sub-goals: sequence, choice and parallelism. These operators allow us to create complex schemes in which the agents can commit to advanced missions.

\paragraph{Deontic Specification:}
The relation between the structural and functional specification is made explicit by the deontic specification. Using it, we can constrain the agents further by specifying what missions an agent \emph{ought} to follow and what missions an agent \emph{is allowed} to follow when playing certain roles. We write
\[
obl(\rho, m, tc),
\]
when agents playing role $\rho$ are obliged to complete mission $m$ under the time constraint $tc$. Analogously we write $per$ for permissions.

The \moise\ organizational model gives a foundation for defining and using an organizational model for multi-agent systems -- in other words to create an OCMAS. However, the model itself is not directly associated with any multi-agent framework and the intention is that it should be usable for all kinds of frameworks for multi-agent systems. 

The software implementation called \jmoise\ is an implementation of \moise\ which should enable multi-agent systems implemented in \jason\ to follow an organizational structure \cite{Hubner+2005}.


\subsection{Implementational details}
\begin{floatingfigure}[lrp]{0.4\textwidth}
\centering
\includegraphics{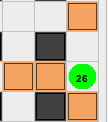}
\caption{The agent is stuck between the boxes.}
\label{fig-acmas-agent-stuck}
\end{floatingfigure}
\noindent
The team-based Bomberman introduced above has been implemented in both an ACMAS and OCMAS version. The following describes some of the implementational details of these systems. Both systems are built for the same environment, and a lot of their abilities will be similar (if not identical). The differences are mostly \emph{how} the systems make use of their knowledge and actions, and not as much \emph{what} abilities they have. This ensures a somewhat identical setup for both systems.

It is the intention that some cooperation is implicitly present. When two agents pursue the goal of killing the same enemy, they should at least be able to avoid putting bombs at the same spots, and instead attempt to trap the enemy. This will generally be possible because of the autonomy of the agents; they should choose paths and bomb locations which seems reasonable, i.e. not place bombs which will potentially hit allies or go through a path in which a bomb may explode soon. 

Some cooperation can, however, not be done implicitly. Consider the situation in figure \ref{fig-acmas-agent-stuck}. Agent ``26'' is stuck between a number of boxes without the possibility of placing a bomb to destroy them; it would kill the agent as well. 

In such situation the agent has two options: (1) wait for another agent to autonomously choose to help the agent or (2) ask for help. Option (1) may be possible but it seems irrational to wait for another agent to detect the situation by himself. The agent being stuck should therefore always ask for help. This is done using the contract net protocol \cite{Wooldridge+2009}.\\
\\
The agents uses the well-known A*-algorithm \cite{Russell+2003} for pathfinding. In addition to the heuristics used by the algorithm, we use a \emph{punishment} value for every location in the environment. The main difference of this algorithm compared to A* is that included in the tentative value of a neighbor of the current location is a \emph{punishment} value depending on the objects on the location of that neighbor. This will make the algorithm consider other, perhaps longer, paths, which however may prove to be safer. 

\begin{figure}[t]
\begin{center}
	\subfigure[A path avoiding boxes.]{
		\includegraphics[height=1.3in]{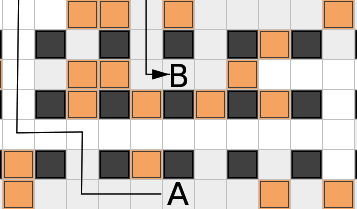}
		\label{fig-acmas-long-deroute-a}
	}
	\subfigure[A path with an intermediate target, $C$.]{
		\includegraphics[height=1.3in]{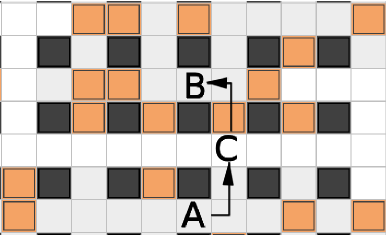}
		\label{fig-acmas-long-deroute-b}
	}
	\caption{Different ways of computing a path from a location $A$ to a target $B$.}
	\label{fig-acmas-long-deroute}
\end{center}
\end{figure}

For instance, by specifying a punishment of 5 on a field containing a box, the algorithm will consider paths, that avoids going through that box, which are up to 5 steps longer. This may not seem as a big improvement, but it means that if a single box is blocking a path, the agent will consider a path which is a little longer. Compared to a situation where he blows up the box and continues, this is usually more efficient, since he will have to wait for the bomb to explode and the explosion to disappear. However, consider the situation depicted in figure \ref{fig-acmas-long-deroute-a}. The agent wants to move from the current position, $A$, to the \emph{target}, $B$, so it will be highly inefficient to compute a path avoiding the boxes. A much more efficient path would be to compute a path, in which a box must be destroyed. To do this we introduce the notion of an \emph{intermediate target}. An intermediate target is a target in which a bomb should be placed in order to clear the way to the ``real'' target. In figure \ref{fig-acmas-long-deroute-b} we have an intermediate target, $C$. This path is clearly preferred over the other. 

The fact that AgentSpeak is Prolog-like makes it possible to specify a model for the plans in a way that is easily transferable to \jason. The agent needs the following predicates from its knowledge base: 
\begin{itemize}
\item $\textsf{pos}(X,Y)$: The current position of the agent.
\item $\textsf{target}(X,Y)$: A final target.
\item $\textsf{intermediate}(X,Y)$: A possible intermediate target.
\item $\textsf{clear}(X,Y)$: An intermediate target is clear, meaning that the targeted box has been removed, creating a passage.
\item $\textsf{bombs}(N)$: The number of bombs currently available. As a shorthand we write $\textsf{bombs}(N_{>0})$ for $(\textsf{bombs}(N) \land N > 0)$.
\end{itemize}

We now present a model for agent $a$ describing the situation above. Note that in a scenario of Bomberman, there will generally be an enormous amount of epistemic states, since there will be a state for each possible position, and an agent will have different knowledge every time he is in a state, yielding even more states. Therefore, we consider a more abstract and general model with three possible locations: $(X_{A},Y_{A})$, $(X_{B},Y_{B})$ and $(X_{C},Y_{C})$, corresponding to the agent's location, his target and intermediate target, respectively. These will be referred to as $A$, $B$ and $C$. We write $\textsf{pred(\us)}$ to match any value of that predicate (i.e. only in the fact that the predicate exists in the knowledge base is of interest).

For each of the three possible locations, a number of possible states exist. In epistemic logic indistinguishable states are typically states the agent cannot distinguish because of lack of knowledge. In the following we also refer to states as indistinguishable if they, even though they are somewhat different in terms of knowledge, will result in the agent performing the same action. This simplifies the model greatly and can be considered as a model created from another agent's point of view. 

\begin{figure}[htbp]
\begin{center}
\includegraphics[width=\textwidth]{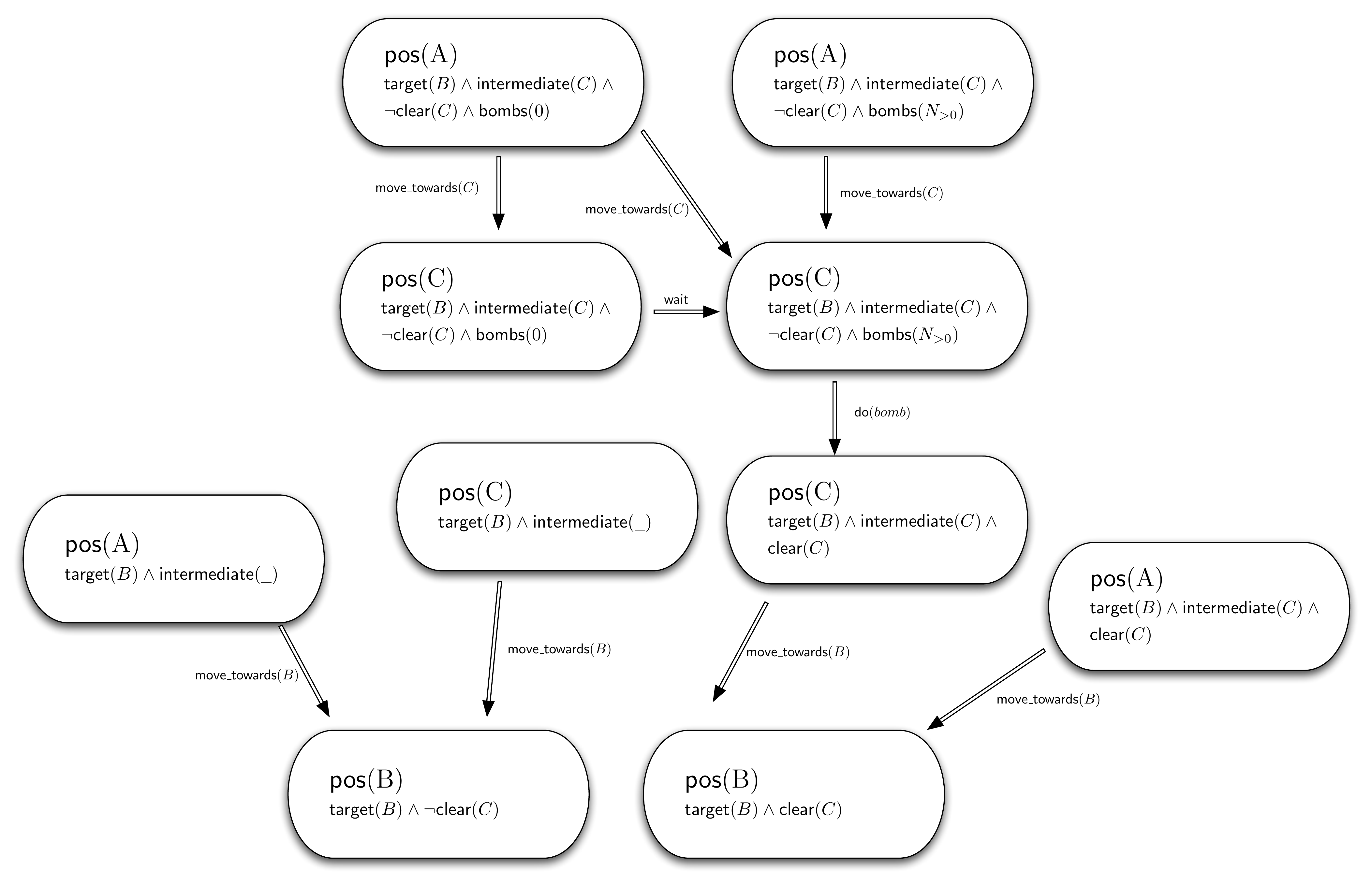}
\caption{Model of the path-finding problem depicted in figure \ref{fig-acmas-long-deroute}.}
\label{fig-acmas-model-pathfinding}
\end{center}
\end{figure}

Figure \ref{fig-acmas-model-pathfinding} shows the model for the path-finding problem. The predicates in a state are the predicates which the agent knows (or believes) to be true at that state. That is, if the agent is at the final target, $B$, and the intermediate target is clear, he will be in the lower right state of the figure. We use this model to create a plan for how to decide which path to choose.

Notice the two possible outcomes of moving towards $C$ when the agent is at $A$ and has no bombs (with the intermediate target being blocked). This is because of the fact that during the move towards $C$, a bomb may become available, meaning that the agent will be in a state where $\textsf{bombs}(N_{> 0})$ rather than $\textsf{bombs}(0)$. 

\paragraph{Committing to a mission in \jmoise}
When the agents commit to a mission in a scheme the \jmoise\ engine will generate goal achievement events for the goals that are currently available. For instance, when an \textsf{explorer} commits itself to the mission of exploration, it will automatically generate the goal achievement event of finding an unexplored area. Whenever a goal is completed, an event for the next goal of the plan is generated. In this case the next available goal will be to move to the unexplored area. In this way, it is very easy to follow the plan of a mission, since the goals are automatically generated when they have been specified in the organization.

Basically, the \textsf{explorer} has the following plans:\\[-20pt]
\begin{verbatimtab}

+!exploreMap[scheme(Sch)]
  <- jmoise.set_goal_state(Sch, exploreMap, satisfied).
			
+!findUnexploredArea[scheme(Sch)]
  :  <context>
  <- <plan to find unexplored area>;
     jmoise.set_goal_state(Sch, findUnexploredArea, satisfied).
			
+!moveToUnexploredArea
  :  <context>
  <- <plan to move to unexplored area>.
	
+near(_,_)
  <- ?scheme(exploration, Sch);
     jmoise.set_goal_state(Sch, moveToUnexploredArea, satisfied).
     
\end{verbatimtab}

Notice that since the organizational specification shows exactly how to explore the map, it is only necessary to create plans for each goal event. When the plan is successfully executed, the agent informs \jmoise\ that the goal has been satisfied. It is then the responsibility of \jmoise\ to generate the next goal event. Note that moving to an unexplored area is a bit different since it uses the path-finding algorithm described in the previous chapter. Therefore, the goal is satisfied only when the agent is near the unexplored area. 

\section{Results}
We now present the main results gained during the work with \jason\ and \jmoise.

\subsection{Agent-Centered Multi-Agent Systems}
The ACMAS, along with its structure, is built form the ground. Having to build everything from the ground gives a lot of freedom with regards to the structure of the implementation; there are no constraints as to where specific details must be implemented. This has lead to a solution where plans for achieving sub-goals and reacting to percepts can be implemented concisely, while still doing as intended. 

The resulting agents are therefore reacting quite fast to changes in the environment; with short code and only few precisely defined responsibilities, the agents are easily able to prioritize during a game, if it, for instance, is necessary to take cover from a bomb. 

But the freedom one has with regards to structure has also been the biggest issue during the implementation. Ensuring successful transition between goals has caused some trouble during the implementation. The debugging functionality can be quite tricky to master, and the only way to test the transition from one goal to another is by executing the system. This can render the process quite slow.

Overall though, we are quite satisfied with the resulting system; it satisfies the proposed strategy and even though the agents may be quite simple, they are able to cooperate to complete their tasks and use their knowledge to decide how to move through the environment.

\subsection{Organization-Centered Multi-Agent Systems}
Building an OCMAS is a more well-structured process than that of building an ACMAS, since it consists of two parts, (1) specifying organization and (2) implementing the details of the organization where the latter depends on the completion of the first. While this does not automatically result in a more structured program, it does force the user to think more about \emph{what}, \emph{why} and \emph{how}. When specifying the organization the focus is on \emph{what} the overall goals are. This leads to considering \emph{why} these are the goals and in that way it allows us to justify the choices made, even before they are implemented. Finally, when the plans are implemented the focus is on \emph{how} the agents are supposed to complete their goals.

Generally, since only sub-goals, and not their relations, need to be implemented, the code tends to be quite clear. However, without being able to study the specification of the organization, it is not possible to see the relation between the goals (as it is handled automatically). Furthermore, as required by \jmoise, the code is often quite verbose, because of the statements required to setup and manage the organizational structure (\texttt{jmoise.create\_group(...)} etc.) and the extensive use of annotations (\texttt{+!goal[scheme(Sch)]} etc.). While this in general makes the code quite clear, it also can result in situations where one need to include a plan for a goal event in which the goal is simply set to be satisfied. Consider the example below:
\begin{quote}
\begin{verbatim}
+!goal // available when subgoal is satisfied
  <- jmoise.set_goal_state(goal, satisfied).
  
+!subgoal
  <- <complete subgoal>;
     jmoise.set_goal_state(subgoal,satisfied).
\end{verbatim}
\end{quote}

In this case, even though the primary goal is completed, when the subgoal is completed, one has to explicitly state that the goal is satisfied even though nothing else happens. This is not a serious problem, but in large scenarios with complicated schemes, it may result in many ``empty'' plans.

Since the agents are part of an organization, they are required to do whatever their obligations tells them to do. To be able to do so, they need access to the specification of the organization. As described, this access is provided through a special agent. This means that when an agent has satisfied a goal, this information is sent to the \jmoise\ agent which then determines what the next goals are and informs the agent. This can decrease performance in very active environments if the agent has no goals to pursue while waiting for new information. 

This can be seen quite clearly in the implementation of the \textsf{explorer} of the OCMAS team. When the \texttt{exploreMap} goal is available, it is immediately satisfied. When this happens, the scheme for exploring is finished and a new scheme must be created in order to continue the exploration. Compared to the \textsf{explorer} of the ACMAS, the performance differences are quite clear; in the ACMAS, the agent immediately chooses a new spot to move to, while in the OCMAS, the agent waits for the generation of an appropriate goal events. 
%
%
%
%
The resulting implementation fulfills the proposed strategy, however the road towards this result has been more bumpy than when working with the ACMAS. This will be discussed below.

\subsection{Using \jason\ and \moise}
\jason\ uses AgentSpeak which is an agent-oriented programming language. Such a programming language is perfect for implementing goal-directed and reactive behavior since one builds a set of plans for how to react to such events. AgentSpeak is very similar to the logic programming language Prolog, and both the language of AgentSpeak and the general features of \jason\ have been quite extensively documented in \cite{Bordini+2007}. This makes it possible to quite easily understand and exploit the features of the interpreter.

The debugging feature of \jason\ has lead to a few issues during the implementation. Often when attempting to debug, the entire system pauses and then when attempting to perform a stepwise operation through the system, nothing happens. This has lead to much trial and error and has overall slowed the development process. Generally though, the system provides descriptive error messages and the more acquainted one gets with the system, the easier errors are spotted.

\textbf{The \jmoise\ extension} is built on \jason\ and uses the \moise\ model, so in general the same things apply to this. However, since it is an extension, it allows for more actions and there are a few more things one needs to be aware of. 

As mentioned, having an organization often leads to a very well structured result since the user is required to really think about what the agents are supposed to do. This is even more the case in \jmoise\ since goal events are automatically generated, meaning that the user need not consider the transition between the goals. Furthermore, the schemes that can be specified in the functional specification of \moise\ makes coordination of tasks very easy. Simply by specifying the cardinality of a goal in a scheme, the user specifies how many agents must complete this goal before it is completed within the scheme. For instance, a goal event can be synchronized by having a sub-goal that all agents must satisfy before the actual goal event is created. 

The \moise\ organizational model has been quite extensively described in \cite{Hubner+2002,Hubner+2005,Hubner+2007} along with a tutorial of the details of how to use it in \cite{Hubner+2008}. This makes it very easy to understand how the different concepts are related and should be used.

\paragraph{Organizational knowledge}
When a group or scheme is created, the agents will perceive certain events so that they are able to react accordingly. In order to be able to distinguish between similar events, annotations are added that among other things include which agent created the organizational object. 

This can be used to let an agent decide not to join a group if a specific agent has created it, or only committing to a mission it is permitted to commit to, if it is related to a specific group. This is a great use of the \jason\ annotations, as it is perfectly clear how to use them. Furthermore, because it is annotations they will not be shown if the programmer chooses not to use them.

What \moise\ is lacking in term of organizational knowledge is the ability for an agent to know whether it is allowed to join a group \emph{before} it attempts to join it. The reason for this is that if it is not permitted to join a group or play a role, an error event is created. This should be okay, but it is not possible for the agent to reason about the error in details so it will not know why it could not join the group. 

Overall, both tools are quite pleasant to work with once acquainted with them.

\section{OCMAS vs ACMAS: When to Use What?}
The work with the two approaches has lead to a discussion of the implementation of each team as well as the two tools used for building the implementations. Generally speaking, one approach is not better than the other but given the results above it is clear that there are situations more suited for one approach than the other. 

\begin{figure}[t]
\begin{center}
\includegraphics[scale=0.6]{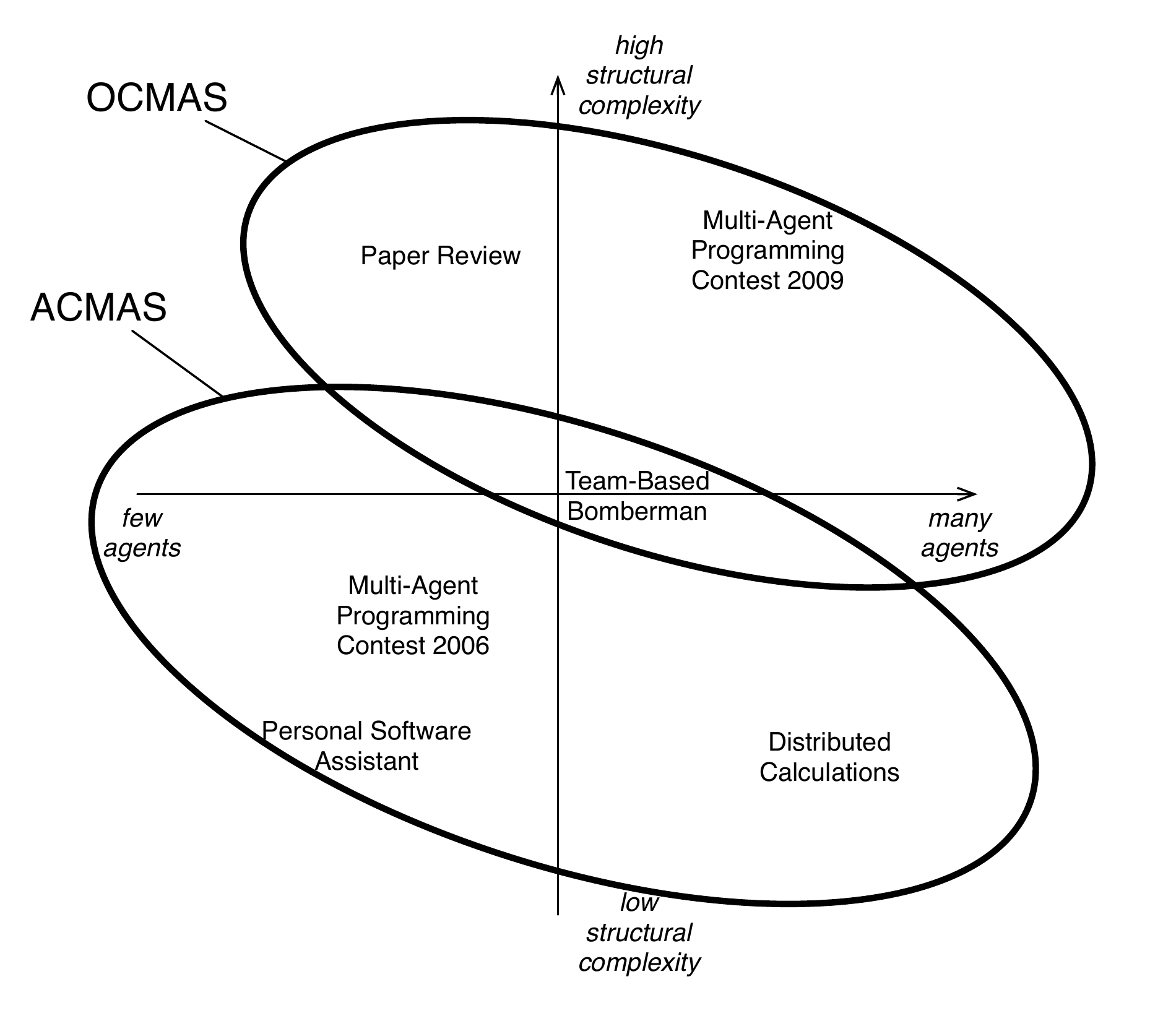}
\caption{An overview of the main results.}
\label{fig-results-overview}
\end{center}
\end{figure}

Figure \ref{fig-results-overview} gives an overview of the main results of the comparison. The figure uses two parameters as basis: the number of agents in the system and the structural complexity. When the system has a high structural complexity, there are greater advantages of dividing the implementation into two distinct parts: \emph{what} and \emph{how}. This means that while an organization can be applied to simple systems, they will in most cases not benefit much from this. 

Notice that the two approaches overlap. The reason is that there will always be situations where it is not clear whether one system is an advantage over the other. The Team-Based Bomberman is an example of such system. While the results show a bias towards the \textsf{ACMAS} in this case, it is partly due to the communication overhead in \jmoise. If this problem is solved, the \textsf{OCMAS} could be performing just as well as the \textsf{ACMAS}.

\paragraph{Personal software assistant:}
A personal software assistant is a simple agent resposible for assisting an end-user with certain tasks that can be automated. For such purpose it seems an organization will be inappropriate. The primary reason for this is that the system will generally consist of very few agents (in many cases only one) and the complexity will be low. In such cases there is not much sense in creating an organization, since the system will not benefit from the OS. It will probably consist of a single group with few roles that the agents can play. Building an entire organization for very few agents can in most cases not be justified. 

\paragraph{Distributed calculations:}
Consider a system of intelligent agents, which have one or more sensors. This could for instance be the ``distributed sensing'' scenario described in \cite{Wooldridge+2009}. Here an agent has a clear responsibility of sensing the environment and using its calculations in some well-defined way. The role of an agent is defined by the sensors it can use, i.e.\ it has a static role. At all time the information it is computing will be used for the same purpose. 

In such cases there is no need to build an organization since the missions are very simple and there is no explicit need of coordination. Furthermore, being in a group would not change the behavior of an agent in the system since its role and responsibilities remains. 

In other words: Even though the system can contain many agents, the structural complexity remains low. Therefore the agents will not benefit much from making the organization explicit.

\paragraph{Paper review process:}
In \cite{Ferber+2004} an example of the ``reviewing process'' of papers in a conference is considered. In this example we have a group for submission of papers and one for evaluation. Being in a certain group then gives an agent certain responsibilities, such as evaluating the papers that are being submitted. 

This is a very specific example of the use of an organization but it can easily be generalized to situations where certain agents are depending on results from other agents. By grouping such agents and creating schemes they can commit to, the general structure of the dependence-relation is immediate and the implementation is easily constructed using it. 

While the number of agents may vary it is clear that the structural complexity is higher than in the previous examples. An explicit organization will definitely make the implementation much easier since responsibilities and acquaintances are well-structured.

\paragraph{Games:}
Games can be quite different and naturally there is no definite answer to whether using \textsf{ACMAS} or \textsf{OCMAS} would be better. In such situations it is important to realize how complex the game is. If the game consists of one well-defined type of controllable character, an organization is probably not a good choice. However, if the game consists of several different characters, all with different possibilities, it may be reasonable at least to consider whether an organization could be useful.

This seems to indicate that using an organization for a Bomberman game may not be the best choice. In this specific case the \textsf{ACMAS} solution is better than the \textsf{OCMAS}; the agents react faster, can more easily adapt to changes and is in general more robust. An organization can be justified in a game such as Bomberman if we include features from the original game that would make the game more complex (e.g.\ power-ups). An \textsf{OCMAS} will benefit in this case, since it is possible to specify advanced roles and missions in the OS that would otherwise be difficult to implement.

\paragraph{Multi-Agent Programming Contest: }
The area of multi-agent systems is quite active, which for instance can be seen by the annual multi-agent programming contest. The primary aim of the competition is to ``stimulate research in the area of multi-agent system development and programming''~\cite{web+mapc}. This is achieved by developing a scenario of a dynamic environment in which cooperation is the key to success. Different multi-agent systems are competing in the scenario in a set of games to determine their performance. As illustrated in figure \ref{fig-results-overview} the complexity and number of agents in the contest has increased over the years. Therefore, while the implementations would not benefit much from an explicit organization in the first scenarios (where only a few agents where required to solve simple tasks), the increased complexity has made this approach a reasonable choice (in scenarios where more than 10 agents are required to cooperate in order to succeed).

The overall complexity of the scenarios in the competition has increased, meaning that it may be easier to implement a better strategy using \textsf{OCMAS} simply because the complexity is easier handled when the structure of the solution has been made explicit.

\section{Future Work \& Conclusions}
By taking both the \emph{agent-centered} and \emph{organization-centered} approach for implementing the same strategy we have gained insights about both the advantages and disadvantages of each approach. The focus has been on a single scenario, which means that not all corners of the approaches have been investigated. Even so, the results have made several differences of the approaches clear, differences that in some situations makes one approach highly advantageous compared to the other.

The focus has been on differences between different types of multi-agent systems and in particular the use of two specific tools. However, it was shown that both types of systems are useful in different situations and that there is no definitive answer to when one system is a better choice than another. The main reason is the many factors the programmer must consider when choosing between the agent- and organization-oriented approaches. These factors include the quality of the implementation, the actual tools used and the strategy to be implemented.

While we have been able to discuss the differences and make suggestions for which system is most suitable in different situations, it would be interesting to be able to create systems of both types which exhibit the same behavior in most situations. This would make it possible to compare the actual performance difference between the systems. However, since this requires specialized systems, there is a chance that the results would not apply to real-world applications.  

The area of multi-agent systems is still somewhat new and is continuously growing. With the addition of the organizational aspects it has been made possible to create even more sophisticated and advanced systems. This comparison has shown that both approaches have advantages and disadvantages and are well-suited for different situations. 

In the end, it is hard to say which approach is better and a decision should be justified by doing some research on the application at hand and the possible tools for creating the system. 

There is still much work to be done in the area of organizational multi-agent systems, specifically in the \moise\ organizational model, but also the principles of OCMAS in general. 

The tools available makes it possible to implement advanced systems (both ACMAS and OCMAS) which are very useful in both research and practical applications and there is no doubt that the area will continue to develop even more efficient and intelligent solutions to research problems and real-world applications. 

\appendix

\section*{Acknowledgement}
I thank associate professor J{\o}rgen Villadsen, DTU Informatics, Denmark.

\end{document}